# Changing the Direction of the Economic and Demographic Research

*By* Ron W. NIELSEN †

**Abstract.** A simple but useful method of reciprocal values is introduced, explained and illustrated. This method simplifies the analysis of hyperbolic distributions, which are causing serious problems in the demographic and economic research. It allows for a unique identification of hyperbolic distributions and for unravelling components of more complicated trajectories. This method is illustrated by a few examples. They show that fundamental postulates of the demographic and economic research are contradicted by data, even by precisely the same data, which are used in this research. The generally accepted postulates are based on the incorrect understanding of hyperbolic distributions, which characterise the historical growth of population and the historical economic growth. In particular, data used, but never analysed, during the formulation of the Unified Growth Theory show that this theory is based on fundamentally incorrect premises and thus is fundamentally defective. Application of this simple method of analysis points to new directions in the demographic and economic research. It suggests simpler interpretations of the mechanism of growth. The concept or the evidence of the past primitive and difficult living conditions, which might be perhaps described as some kind of stagnation, is not questioned or disputed. It is only demonstrated that trajectories of the past economic growth and of the growth of population were not reflecting any form of stagnation and thus that they were not shaped by these primitive and difficult living conditions. The concept or evidence of an explosion in technology, medicine, education and in the improved living conditions is not questioned or disputed. It is only demonstrated that this possible explosion is not reflected in the trajectories of the economic growth and of the growth of population.

## 1. Introduction

What we are going to see will change the fundamental postulates in the demographic and economic research. It will change radically the way the mechanism of economic growth and of the growth of population is interpreted. Maybe the change will not come immediately because it is usually difficult to change the well-established interpretations and explanations but the change will come because this is the way science works. Incorrect interpretations are not tolerated for too long and it does not matter who are their advocates.

It might be expected that a complicated proof would be required to achieve such a radical change of direction in the economic and demographic research, that perhaps some new and complicated description of the mechanism of growth would have to be proposed. However, the proof turns out to be exceptionally simple. No complicated mathematics is required but only the way we describe data using the simplest mathematical representation: the straight line.

George Pólya, Hungarian mathematician, observed that when a proof is too simple, "youngsters" will be unimpressed (Pólya, 1981), but mathematics does not have to be complicated to be useful. He also pointed out that solving problems is a quintessential human activity and the aim is always to find *the simplest solutions*.

We are going to present here a proof so simple tit might look trivial. We are going to show how to change the confusing and complicated distributions describing the historical economic growth and the historical growth of human population into the simplest representations. We are going to show how the distributions, which suggest complicated explanations of the mechanism of growth are in fact so simple that they suggest also a simple mechanism.

---

† AKA Jan Nurzynski, Griffith University, Environmental Futures Research Institute, Gold Coast Campus, Qld, 4222, Australia.

☎. +61407201175

✉. ronwnielsen@gmail.com

Analysis of data describing the historical economic growth and the historical growth of population might look complicated but it is exceptionally simple. Anyone can do it. However, there is more to the analysis of data then just looking for their mathematical descriptions. We are going to demonstrate that this simple method of analysis makes a significant contribution to a better understanding of the mechanism of the historical growth of population and of the economic growth. It also demonstrates that there is a need to replace the traditionally used postulates based largely on impressions and conjectures by postulates based on the mathematical analysis of data.

## 2. The common problem

Hyperbolic processes appear to be causing a serious problem in the economic and demographic research. They create such a strong illusion that it deceives even the most experienced and respected researchers. The common mistake is to see them as being made of two distinctly different components, slow and fast, with a clear transition between them (Ashraf, 2009; Artzrouni & Komlos, 1985; Baldwin, Martin & Ottaviano, 2001; Becker, Cinnirella & Woessmann, 2010; Clark 2003, 2005; Currais, Rivera & Rungo 2009; Dalton, Coats & Asrabadi, 2005; Desment & Parente, 2012; Doepke, 2004; Ehrlich, 1998; Elgin, 2012; Galor 2005a, 2005b, 2007, 2008a, 2008b, 2010, 2011, 2012a, 2012b; Galor & Michalopoulos, 2012; Galor & Moav 2001, 2002; Galor & Mountford, 2003, 2006, 2008; Galor & Weil, 1999, 2000; Goodfriend & McDermott 1995; Hansen & Prescott 2002; Jones, 2001; Johnson & Brook 2011; Kelly, 2001; Khan 2008; Klasen & Nestmann 2006; Kögel & Prskawetz 2001; Komlos 1989, 2000, 2003; Komlos & Artzrouni 1990; Lagerlöf 2003a, 2003b, 2006, 2010; Lee, 2003, 2011; Mataré, 2009; McFalls, 2007; McKeown, 2009; McNeill 2000; Møller & Sharp, 2013; Mongomery, n.d.; Nelson, 1956; Omran 1971, 1983, 1986, 1998, 2005; Robine 2001; Smil 1999; Snowdon & Galor, 2008; Steinmann, Prskawetz & Feichtinger, 1998; Strulik, 1997; Tamura 2002; Thomlinson 1965; van de Kaa 2008; Voigtländer & Voth, 2005; Vollrath, 2011; Wang 2005, Warf 2010; Weisdorf 2004; Weiss 2007). The next step is then to try to explain these two perceived stages of growth and the associated but non-existent transition by proposing distinctly different mechanisms for each of these imagined components rather than seeing them as representing a *single*, *monotonically increasing* distribution governed by a *single mechanism of growth*.

This step leads progressively further away from the correct understanding of studied processes because all efforts are now concentrated on explaining the non-existing features. An increasing number of scholars are being involved. They do not analyse the relevant data but only describe their impressions created by hyperbolic illusions. The participating researchers do not question the existence of the distinctly different stages of growth or of the postulated transition – they take them for granted and concentrate their attention only on the explanation of these phantom features, proposing new mechanisms, theories and mathematical descriptions without realizing that the apparent distinctly different two stages of growth do not exist and that there is no transition but a monotonically increasing hyperbolic distribution. Their mathematical descriptions, complicated and elaborate as they might be, are not the descriptions of the studied processes but rather the descriptions of phantom impressions created by hyperbolic illusions.

The perceived two stages of growth are commonly described as stagnation and sustained growth, while the perceived but non-existent transition as an escape, sprint, sudden spurt, intensification, acceleration, explosion or by some other similar terms all emphasizing a clear and dramatic change in the pattern of growth at a certain time. Variety of forces and mechanisms are then proposed to explain the phantom stages of growth and of the associated but non-existent transition. Efforts are also made to determine the precise time of the non-existent transition, often placing it around the Industrial Revolution but sometimes around 1950, without realizing that the determination of this time is impossible because there was no unusual acceleration at any particular time or over a certain range of time.

Hyperbolic processes are prone to misinterpretations and consequently they have to be analysed with care. Fortunately, their analysis is exceptionally simple. To show how to avoid being guided by hyperbolic illusions we shall describe the simple method of their analysis and illustrate it by a few examples.

## 2. The method of reciprocal values

Hyperbolic processes can be easily analysed using the method of reciprocal values. This method is so simple that it can be explained by using just two elementary equations, and yet so powerful that it can turn around and revolutionize such fields of research as the economic growth and the growth of human population, the important fields of study because for the first time in human existence we have now reached ecological limits of our planet and the correct understanding of these two processes is essential to avoid the undesirable unsustainable developments. We have to know how these processes work and how to control them. Incorrect interpretations are potentially dangerous and cannot be tolerated. Every effort has to be made to identify and eliminate any incorrect and misleading explanations.

The first-order hyperbolic distribution is described by the following simple equation:

$$S(t) = (a_0 + a_1 t)^{-1}, \qquad (1)$$

where $S(t)$ is the size of a growing entity, while $a_0$ and $a_1$ are constants. For the hyperbolic growth, $a_1 < 0$.

Example of hyperbolic growth is shown in Figure 1. It represents the growth of the world population during the AD era. We can see that hyperbolic distribution describes well the growth of population during the entire range of data.

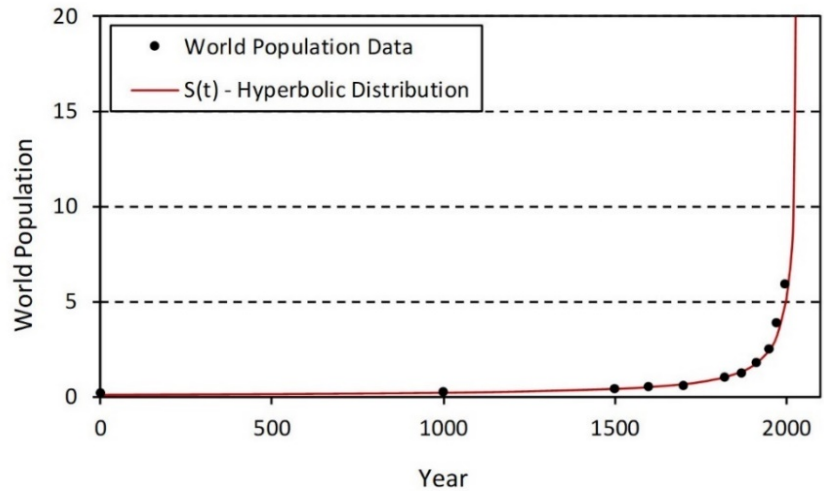

**Figure 1.** Example of hyperbolic growth. Population data (Maddison, 2001) taken from the same source as used by Galor in his Unified Growth Theory (Galor, 2005a, 2011) are compared with hyperbolic distribution.

Data have to be analysed but in general they are not. Meticulous analysis of data is particularly important in the study of hyperbolic processes because they may be strongly misleading. They easily create an illusion of stagnation followed by explosion. Unfortunately, on seldom occasions when data are used and displayed, they are displayed in a grossly distorted and self-misleading way (Ashraf, 2009; Galor 2005a, 2005b, 2007, 2008a, 2008b, 2010, 2011, 2012a, 2012b; Galor & Moav, 2002; Snowdon & Galor, 2008) as shown in Figure 2.

Figure 2 was repreduced from Galor's publication (Galor, 2005a, p. 181). His figure was based on precisely the same source of data (Maddison, 2001) as used in Figure 1 but in this distorted way they show no resemblance to the the original data. Such distortions were used repeatedly during the development of the Unified Growth Theory (Galor, 2005a, 2011) making it scientifically unacceptable, incorrect and unreliable. This Figure shows incorrectly that there was a long epoch of stagnation followed by a takeoff to a fast growth.

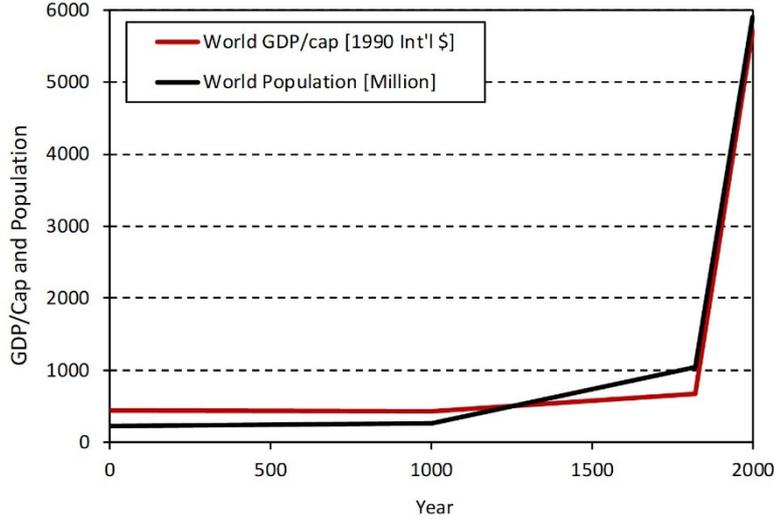

**Figure 2.** Example of a distorted presentation of data used in academic literature (Ashraf, 2009; Galor 2005a, 2005b, 2007, 2008a, 2008b, 2010, 2011, 2012a, 2012b; Galor & Moav, 2002; Snowdon & Galor, 2008). Data presented in this figure come from precisely the same source (Maddison, 2001) as the data presented in Figure 1 but in this distorted way they bear no resemblance to the original data and they suggest incorrect interpretation of the mechanism of growth. Population is in millions and the GDP/cap in the 1990 International Geary-Khamis dollars

In discussions of the growth of population or of the economic growth it is easy to use some selected numbers and show that the growth was slow over a long time and fast over a short time. The slow growth is then interpreted as stagnation controlled by random forces of growth usually associated with Malthusian positive checks. The fast growth is intereprted as explosion controlled by distinctly different forces. The triggering mechanism of the alleged explosion is usually claimed to have been associated with the Industrial Revolution and Galor conveniently locates this alleged takeoff time around the time of the Industrial Revolution. Of course the growth was slow over a long time and fast over a short time because it was hyperbolic. It was not because there was stagnation followed by a takeoff or explosion leading to a new type of growth.

Hyperbolic distribution shown in Figure 1 is described by the eqn (1) with the following parameters $a_0 = 8.724$ and $a_1 = -4.267 \times 10^{-3}$. The fit to the data is remarkably good. Details of analysis are described in a separate publication (Nielsen, 2016a). They show that there was a major transition from a fast hyperbolic growth to a slow hyperbolic growth around AD 1 and that there was a minor disturbance around AD 1300. However, these details are of no concern to us in our present discussion. What is important to notice is that the growth of human population was indeed slow over a long time and fast over a short time but that these features are described remarkably well by a *single* hyperbolic distribution. These features represent nothing more than mathematical properties of hyperbolic distribution. They represent a *single* mechanism of growth.

It is important to point out that hyperbolic distribution increases monotonically. It makes no sense to divide it into two or three components and assign different mechanisms of growth to each perceived component. Hyperbolic distribution cannot and should not be divided into separate components and the best way to see it is to plot their reciprocal values $[S(t)]^{-1}$ because they convert hyperbolic distribution to a straight line:

$$[S(t)]^{-1} = a_0 + a_1 t \qquad (2)$$

Reciprocal values of hyperbolic distribution shown in Figure 1 are plotted in Figure 3. It is precisely the same distribution as shown in Figure 1 but it is presented in a different way. The confusing features such as the apparent stagnation followed by a takeoff to a fast growth increasing to infinity are replace by a clear straight line, which is easy to understand. It is obvious now that it would make no sense to divide such a straight line into distinctly different components and to claim distinctly different mechanisms of growth. It is also clear that it is impossible to identify a transition from a slow to a fast growth for hyperbolic distributions. There is no transition at any time. The transition occurs gradually over the entire range of growth. It is impossible to identify a takeoff time because there was no takeoff.

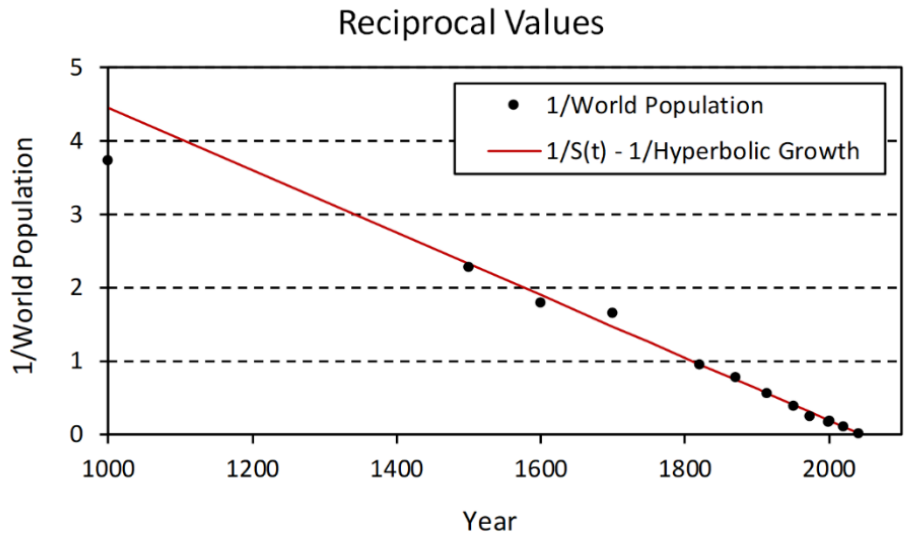

**Figure 3.** Reciprocal values of the hyperbolic distribution presented in Figure 1 together with the reciprocal values of the size of population. Complicated hyperbolic distribution is now represented by a simple straight line, which is easy to understand. The takeoff around 1800 shown in Figure 2 did not happen. The straight line cannot be divided into two distinctly different components making it clear that hyperbolic distribution shown in Figure 1 is also made of a single component. The slow and the fast growth shown in Figure 1 follow the same, monotonically-increasing distribution.

The display in Figure 3 is from AD 1000 for two reasons. (1) There is a large gap between AD 1 and 1000 so the display from AD 1000 shows better the agreement of the fitted hyperbolic distribution with data. (2) Detailed analysis of data for the AD and BC eras shows clearly that between around 500 BC and AD 500 there was a massive transition from a fast hyperbolic growth during the BC era to a significantly slower hyperbolic growth during the AD era (Nielsen, 2016a). The point at AD 1 is right in the middle of this transition and belongs to an entirely different distribution, the distribution describing the process of transition.

It should be also noticed that the point at AD 1000 in Figure 3 appears to be much further away from the fitted distribution then the point in Figure 1. The distributions are precisely the same but the display of reciprocal values magnifies the discrepancies between data and the calculated curve for small values (Nielsen, 2016b). The smaller are the values of the data and of the calculated distribution the larger is the magnification.

Reciprocal values allow for a unique identification of the first-order hyperbolic distributions because only these distributions are represented then by straight lines. This representation allows also for an easy study of departures from hyperbolic growth because deviations from a straight line are easy to notice.

Properties of growth do not change by changing the display of data but certain features, which are difficult or even impossible to recognize in one display can be easily identified in another. It is essential to remember that in the display of reciprocal values effects are reversed. Thus, for instance, a deviation to a slower trajectory will be indicated by an *upward* bending and deviation to a faster trajectory by a *downward* bending. An increasing growth is represented by a decreasing trajectory of the reciprocal values.

When hyperbolic growth is represented by a mathematically generated and gradually changing curve, such as shown in Figure 1, it might be clear that there was no particular time when the growth changed from being nearly horizontal to nearly vertical, but when data represented by discrete points are displayed, such a conclusion might be less obvious. The illusion becomes particularly strong when only a few strategically located points are selected (Ashraf, 2009; Galor 2005a, 2005b, 2007, 2008a, 2008b, 2010, 2011, 2012a, 2012b; Galor & Moav, 2002; Snowdon & Galor, 2008) from a significantly larger set of data as if to make the deception even more pronounced. Even if the enforcement of the perceived illusion is unintended, such crude displays of data lead readily to grossly incorrect interpretations.

However, if reciprocal values of data are displayed, their analysis is immediately made significantly simpler because if the data follow a simple, first-order hyperbolic distribution, their reciprocal values will be clearly aligned along a decreasing straight line. It is then obvious that dividing such a straight line into two sections and claiming two distinctly different regimes of growth governed by two distinctly different mechanisms simply makes no sense. It also makes no sense to try to locate a point on the decreasing straight line and claim a transition to a new trajectory because there is obviously no transition to a new trajectory on a decreasing straight line.

It should be stressed that in this representation only the first-order hyperbolic distributions describing growth will follow the decreasing straight-line trajectories. It is for this reason that this simple method is so useful in identifying the first-order hyperbolic distributions. It is a simple and yet powerful method, which can be used successfully in the analysis of data describing the historical economic growth and the growth of human population, global, regional or local, because in general they follow simple, first-order hyperbolic trajectories. Any deviations from such trajectories can be easily investigated. Higher-order hyperbolic distributions describing growth will be represented by gradually decreasing trajectories, which could be fitted using higher-order polynomial functions intercepting the horizontal axis, while the exponential growth will be represented by a decreasing exponential function.

This method might have a more general application but its specifically intended application described in this publication is to help to avoid being guided by hyperbolic illusions, the unfortunate common mistake, which often leads to seriously incorrect conclusions as we shall demonstrate in the examples 2 and 3.

Going beyond the intended application, the first-order decreasing hyperbolic distributions will be represented by the increasing straight lines. Again, in this representation, any deviation from the decreasing hyperbolic distributions can be easily detected and investigated. Pareto distributions, which resemble the decreasing hyperbolic distributions, will be represented by gradually increasing functions, which in this representation might be also easier to investigate.

We shall now illustrate the application of the method of reciprocal values by using three additional examples: the growth of human population in Africa, the economic growth in Western Europe and the world economic growth.

## 3. Further examples
### 3.1. Growth of population in Africa

The method of reciprocal values can be used to study fine details of growth trajectories, the study which can then be used not only to improve the fit to data but also to understand the mechanism of growth. Some distributions might be made of different components, which could be difficult or even impossible to see in the direct display of data but they could be easily revealed by displaying their reciprocal values. An excellent example is the growth of human population in Africa shown in Figure 4, constructed using Maddison's data (Maddison, 2010). These Figure illustrates the added advantage of using the reciprocal values of data.

The top panel in Figure 4 contains the direct display of data for the growth of human population in Africa. The displayed shape suggests hyperbolic growth because it is slow over a long time and fast over a short time.

However, the reciprocal values of data presented in the lower panel reveal that the growth trajectory is in fact made of two major components: a slow hyperbolic distribution until around 1870 and a fast hyperbolic distribution after that year. Parameters describing the two hyperbolic components are $a_0 = 5.105 \times 10^1$, $a_1 = -2.036 \times 10^{-2}$ for the slow component and $a_0 = 1.705 \times 10^2$, $a_1 = -8.515 \times 10^{-2}$ for the fast component.

Figure 4 shows also that at a later stage, the fast hyperbolic growth started to be diverted to a slower trajectory as indicated by the upward bending of the trajectory representing the reciprocal values. Furthermore, it is now clear that the growth of population in Africa was never stagnant and that there was never a transition from stagnation to growth. The first stage of growth was hyperbolic and the transition around 1870 was a transition from hyperbolic growth to another hyperbolic growth. All these features, which are unrecognisable in the direct display of data are clearly seen in the display of the reciprocal values.

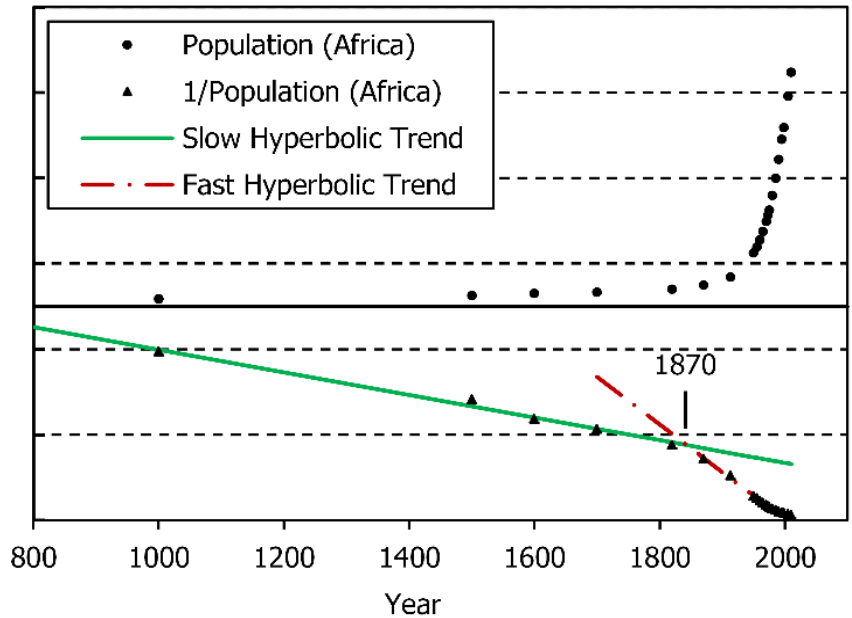

**Figure 4.** Growth of human population in Africa (Maddison, 2010) illustrates how the method of reciprocal values can serve as an excellent tool in revealing hidden features of studied distributions.

The pattern revealed by data contradicts the traditional interpretations of the mechanism growth of human population. There was no escape from the Malthusian trap because there was obviously no trap in the growth of population. The growth was slow but it was increasing monotonically with no signs of restrictions imposed by a mythical trap.

The transition from a slow to a fast hyperbolic growth in Africa occurred around the time of the Industrial Revolution but it was not a transition from a stagnant growth to a new, so-called sustained growth regime (Galor, 2005a, 2011) but from a hyperbolic growth to another but faster hyperbolic growth. It was the boosting that coincides with the intensified colonisation of Africa (Duignan & Gunn, 1973; McKay, Hill, Buckler, Ebrey, Beck, Crowston, & Wiesner-Hanks, 2012; Pakenham, 1992).

Contrary to the commonly accepted interpretations, the boosting in the growth of population was *not* triggered by a dramatically *decreased* intensity of Malthusian positive checks but by their dramatic *escalation*. It is clear that the accepted interpretations of the effects of Malthusian positive checks are incorrect. Their increased intensity does not lead to stagnation but to a more intensified growth (Malthus, 1798; Nielsen, 2016c). The increased intensity of Malthusian positive checks increases the mortality rate but it also increases the fertility rate with the net result of increasing the rate of natural increase or the growth rate. This correlation is also clearly demonstrated even now by the growth of population in poor countries. The poorer they are the faster is the growth of their populations. Thus, this simple analysis of data assisted by using the reciprocal values already questions the commonly accepted interpretations of the mechanism of growth of human population.

As shown in Figure 4, reciprocal values of data reveal the details of the mechanism of growth, which were impossible to identify by the direct display of data. Even if we cannot yet fully explain these details, we can already see that the growth of the populations in Africa was following a slow hyperbolic trend until around 1870. Around that year, the growth of human population in Africa experienced an unprecedented 4-fold acceleration, which diverted the growth into a significantly faster hyperbolic trajectory. The fast-hyperbolic growth continued until around 1975 when it started to be diverted to a new but slower trend.

It is this pattern of growth that we have to explain. It is for this pattern of growth that we have to propose the mechanism of growth. It is not the imaginary pattern of stagnation followed by explosion. It is not the fictitious Malthusian regime followed by the mythical takeoff from stagnation to an imagined sustained growth regime (Galor, 2005a, 2011). It is an entirely different pattern, the pattern indicated by the close analysis of data rather than by the pure fantasy. The aim of scientific investigation is not to explain figments of imagination but the evidence presented by data.

Data are essential in scientific investigations. Assisted by data we shall not be guided by the erroneous concept of stagnation but by the clear evidence of hyperbolic growth. We shall also not be guided by the erroneous concept of a takeoff from stagnation to a sustained growth regime but by the clear evidence of a transition from a hyperbolic growth to another hyperbolic growth. We shall also be guided by an observation that at a certain stage, around 1975, the long-lasting pattern of hyperbolic growth has been eventually abandoned and the growth was diverted to an entirely different trajectory.

### *3.2. Economic growth in Western Europe*

Economic growth is measured using the Gross Domestic Product (GDP) or the GDP per capita (GDP/cap). Galor and Moav (2002) studied economic growth in Western Europe using the data of Maddison (Maddison, 2001). They have selected a few, strategically located points from a larger set of data, joined them by straight lines and concluded that there were two distinctly different regimes of growth: the "Malthusian regime" (also labelled as the "epoch of stagnation," "Malthusian era," "Malthusian epoch," "Malthusian steady-state equilibrium," "Malthusian stagnation" or "Malthusian trap") and the "sustained economic growth" (described also as the "Modern Growth Regime," "sustained economic growth" and "sustained growth regime"). Their distorted representation of Maddison's data is shown in Figure 5.

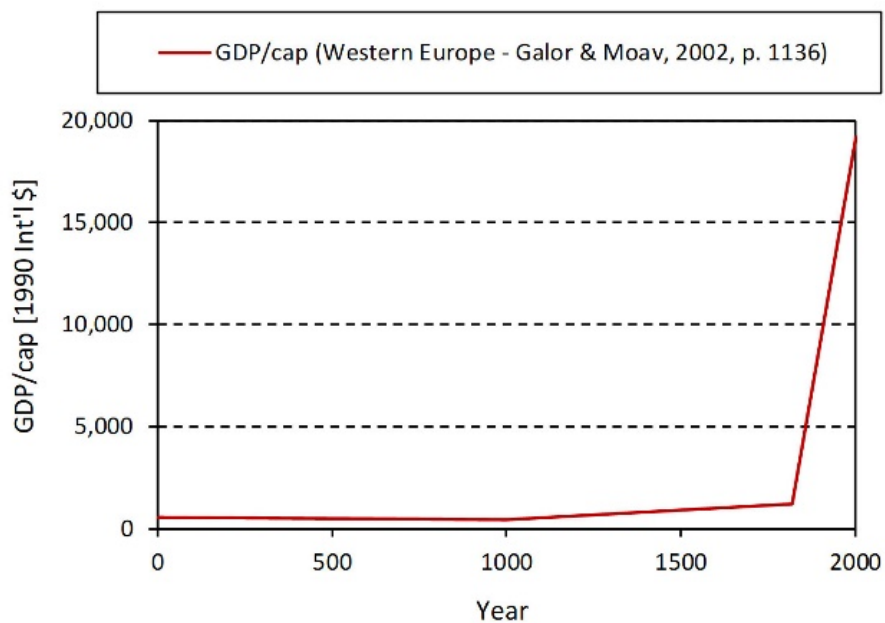

**Figure 5.** A typically distorted and self-misleading representation (Galor & Moav, 2002, p. 1136) of Maddison's data (Maddison, 2001). Compare it with exactly the same data, but not distorted, presented in Figure 7.

Referring to this crude display of data they also concluded that the Industrial Revolution had a strong impact on the economic growth causing a dramatic takeoff from stagnation to a fast growth. They made no attempt to analyse mathematically Maddison's data (Maddison, 2001) but presented a series of mathematical equations describing their imaginations, which were neither related to nor supported by the source of data they have used.

It is remarkable that data coming from *precisely the same source* as they have used contradict their claims and their interpretations of growth. Extensive analysis of the GDP/cap data, global and regional, is presented in a separate publication (Nielsen, 2016d). It is shown there that GDP/cap data follow the *monotonically increasing* trajectories. They are just the linearly modulated hyperbolic trajectories (Nielsen, 2017a), i.e. hyperbolic trajectories modulated by the linear time-dependence of the reciprocal values of the size of population. There is no stagnation and no takeoff to a distinctly different regime of growth. Both, the GDP and the population increase hyperbolically (Nielsen, 2016b, 2016e, 2016f) and thus monotonically. Consequently, their ratios increase also monotonically.

Figure 6 presents the reciprocal values of the Gross Domestic Product (GDP) for Western Europe (Maddison, 2001) in the vicinity of the alleged takeoff. The data are well aligned along a decreasing straight line, which means that they were following the simplest, first-order, hyperbolic distribution given by the eqn (1).

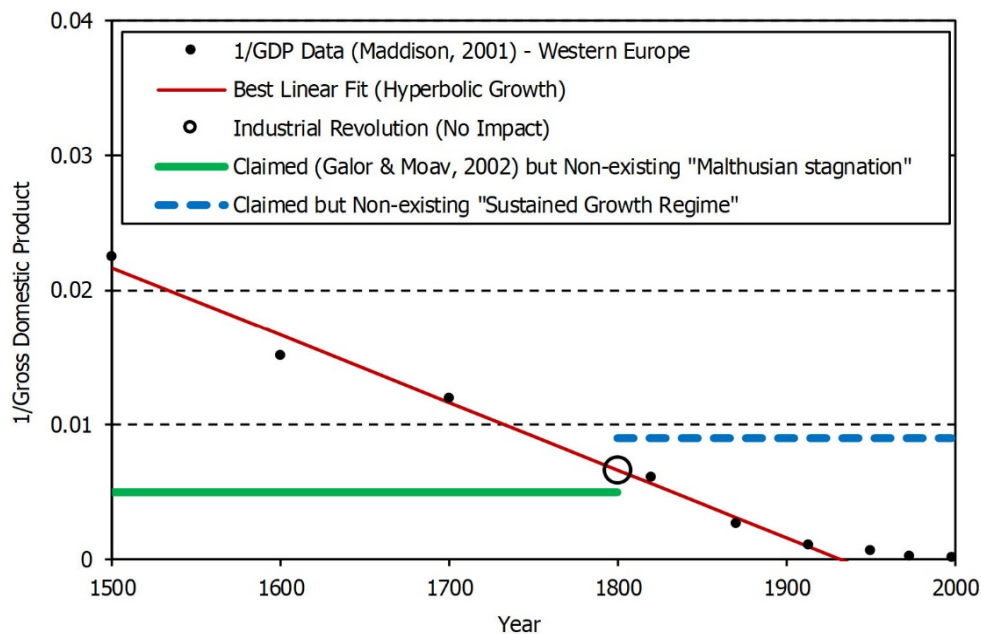

**Figure 6.** Reciprocal values of data describing the Gross Domestic Product (GDP) in Western Europe (Maddison, 2001) in the vicinity of the Industrial Revolution. This is precisely the same source of data as used by Galor and Moav (2002) to construct their distorted representation shown in Figure 5. Contrary to their claim, Industrial Revolution had no effect on shaping the economic growth trajectory in Western Europe, the centre of this revolution. The two regimes of growth claimed by them also did not exist. The GDP is in billions of the 1990 International Geary-Khamis dollars.

Industrial Revolution was between 1760 and 1840 (Floud & McCloskey, 1994), or around 1800 as shown in Figure 6. This figure demonstrates clearly and convincingly that the claimed takeoff around the time of the Industrial Revolution did not happen because the reciprocal values of the GDP data follow an undisturbed straight-line trajectory representing an undisturbed hyperbolic growth. It is now clear that there was no takeoff and no escape, great or small, from the hypothetical but non-existing Malthusian trap, at least from the alleged trap in the economic growth. Maybe there were some other traps but maybe they are just figments of imagination. It is clear, however, that Industrial Revolution had absolutely no impact on shaping the economic growth trajectory in Western Europe, the centre of this revolution.

Industrial Revolution had, no doubt, many other impacts but they are not reflected in the economic growth trajectory. Their study could be important and interesting but they will not explain the growth of the GDP. The mechanism of growth was immune to the changes introduced by the Industrial Revolution. Whatever dramatic changes the Industrial Revolution might have introduced to the general style of living, to technology and even to the economic marked, these changes obviously were not shaping the economic growth trajectory.

The absence of a takeoff eliminates also the need for assuming the existence of two distinctly different regimes of growth. It obviously makes no sense to divide the straight line into two arbitrarily selected sections and claim distinctly different trajectories governed by distinctly different mechanisms of growth. What might not have been clear in the direct display of data, is now perfectly obvious if we display the reciprocal values of data. This display abolishes all elaborate theories and untidy explanations incorporating such concepts as traps, escapes, takeoffs and stagnation and replaces them by a simple interpretation of the mechanism of growth suggested by the simple equation describing hyperbolic growth. This conclusion is in agreement with the general observation that natural phenomena can be usually explained by using simple descriptions.

In Figure 7, the hyperbolic trajectory corresponding to the straight line shown in Figure 6 is extended to AD 1. The economic growth in Western Europe is well described by a simple, first-order, hyperbolic distribution. The corresponding parameters are: $a_0 = 9.697 \times 10^{-2}$ and $a_1 = -5.020 \times 10^{-5}$. The point at 1950 is not fitted by the hyperbolic trend because from the early 1900s the economic growth in Western Europe started to be diverted to a *slower* trajectory, which is again contrary to the claimed boosting or a transition from stagnation to growth. There was a transition but it was a transition from a *monotonically increasing* hyperbolic growth to a *slower* trajectory.

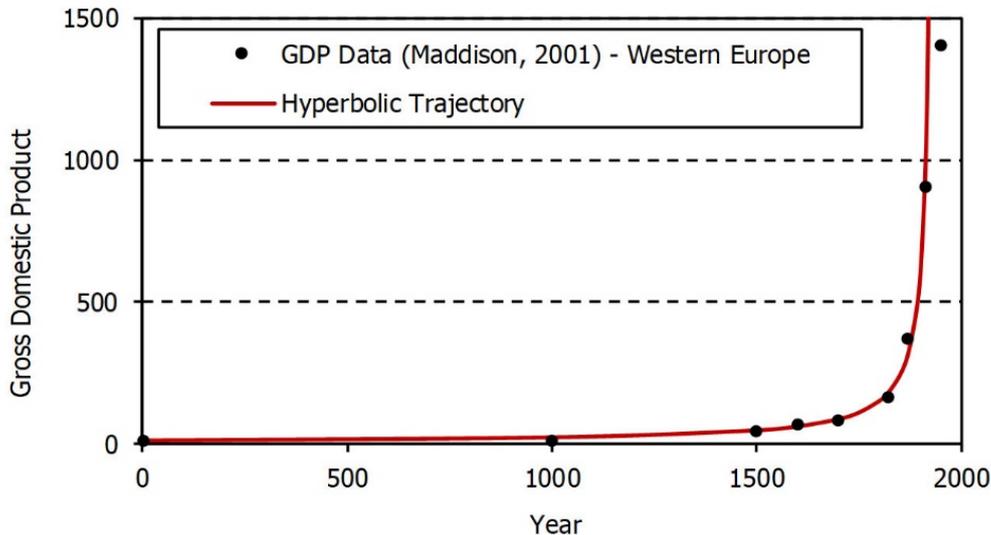

**Figure 7.** Data for the Gross Domestic Product (GDP) in Western Europe (Maddison, 2001) compared with the monotonically increasing hyperbolic distribution. The GDP is in billions of the 1990 International Geary-Khamis dollars.

We cannot claim that the growth was sustained only after the Industrial Revolution because it was sustained equally strongly during the postulated but non-existent "epoch of stagnation." Figure 6 and Figure 7 show clearly that the concept of two stages of growth is unsupported by data. When stripped of the hyperbolic illusions, the economic growth is revealed as a simple process, which can be described using just one, simple mathematical trajectory until the early 1900s when it started to be diverted to a slower, non-hyperbolic, trajectory. There is no compelling need to make this simple description complicated.

Growth of the GDP was slow in the past because it was hyperbolic. However, while being slow it was not stagnant. The growth was fast in recent years because it was hyperbolic. It followed the same undisturbed hyperbolic distribution as in the past.

We now have a completely different understanding of the economic growth in Western Europe, an important turnaround in the economic research. Rather than wasting the valuable time, energy and financial resources on trying to explain the phantom features created by hyperbolic illusions and magnified by the customary crude representation of data (Ashraf, 2009; Galor 2005a, 2005b, 2007, 2008a, 2008b, 2010, 2011, 2012a, 2012b; Galor & Moav, 2002; Snowdon & Galor, 2008) we can now focus our attention on the relevant task of trying to explain why the economic growth was so stable over such a long time and why it was hyperbolic. Rather than writing numerous articles based on impressions and publishing them in peer-reviewed scientific journals and in academic books we can now concentrate our attention on the understanding of *the science* of economic growth. In our investigations, we shall not be guided by impressions, we shall not be guided by the customary crude representations of data but by their rigorous mathematical analysis

## *3.3. Global economic growth*

Another example of the application of the method of reciprocal values is the global economic growth. It is an important example because it questions Galor's Unified Growth Theory (Galor, 2005a, 2011) representing the culmination of his work extending over 20 years (Baum, 2011). His theory is based on an uncritical acceptance of the common interpretations, descriptions and explanations used in the economic and demographic research. In this sense, his theory offers no new insights.

The fundamental postulate of this theory is again the existence of three regimes of growth: the slow and stagnant Malthusian Regime, the short and intermediary Post-Malthusian Regime and the fast, Sustained Growth Regime. Galor also accepts that Industrial Revolution played a crucial role in the alleged dramatic takeoff from a prolonged stagnation into a rapid and sustained growth.

The welcome initiative in his theory is that he makes an attempt of using repeatedly Maddison's data (Maddison, 2001). However, he makes not even a single attempt to test his theory by the rigorous analysis of data. This is a serious omission. The usual practice in any scientific theory is to test it by data or at least to suggest how it can be tested by data. Galor does not follow this accepted practice. He does not test his mathematical descriptions by data. Data are used repeatedly but they are never analysed. They are presented in a typically distorted way, as illustrated in Figures 2 and 5, and in this distorted way they seem to support the preconceived ideas. His work is based on prejudice and no attempt is made to check its validity.

When data are used but manipulated to confirm preconceived ideas we are not dealing with science. We also make no progress and we are not learning anything new or useful.

We shall now use exactly *the same source of data* and show that the Unified Growth Theory is scientifically unsustainable. For more extensive discussion of these issues see other publications (Nielsen, 2016a, 2016b, 2016d, 2016e, 2016f, 2016g, 2016h, 2016i, 2016j, 2017a).

It is hard to see how much can be rescued from Galor's Unified Growth Theory. It is hard to see how many of his descriptions and explanations are based on pure and unsubstantiated speculations. His theory would have to be minutely analysed. However, its major premises are untenable. All his "mind boggling" "mysteries of the growth process" (Galor, 2005a, p 220), for instance, can be easily explained (Nielsen, 2016a, 2016d, 2016g, 2016i) – *there are no mysteries*. All his mysteries were created by his repeatedly distorted presentations of data coming from a reputable source (Maddison, 2001), the data used during the formulation of his theory but never properly analysed.

His theory certainly does not explain the mechanism of growth because it revolves around the descriptions of hyperbolic illusions. It does not even describe economic growth. His descriptions are incorrect because again they are based on the distorted presentations of data and on the unsubstantiated prejudice.

Theories come and go. Scientific integrity is not tarnished by proposing incorrect explanations and interpretations but by refusing to correct them or to reject them when they are contradicted by reliable data.

Reciprocal values of data for the world Gross Domestic Product (GDP) (Maddison, 2001) are shown in Figure 8. They follow closely a decreasing straight line, which means that the economic growth was increasing hyperbolically. It is clear that there was no takeoff of any kind, large or small, around the time of the Industrial Revolution and no repeatedly claimed great escape from the postulated but non-existing Malthusian trap. The data do not support the existence of the three regimes of growth and thus contradict the fundamental postulates of the Unified Growth Theory.

The last point of the data shown in Figure 8 is not fitted by the straight line, suggesting a possible diversion to a slower trajectory. This region can be studied more closely using the extended compilation of the economic growth data (Maddison, 2010). Their reciprocal values between 1700 and 2003 are shown in Figure 9 demonstrating clearly that while the Unified Growth Theory claims an unusually accelerated growth after the alleged but non-existent epoch of stagnation, the data show the opposite behaviour: a diversion to a *slower* trajectory after the earlier vigorous, well-sustained and secure economic growth. Rather than being boosted by the Industrial Revolution, the economic growth continued along the *undisturbed* hyperbolic trajectory for about one hundred years after this revolution and then started to be diverted to a *slower* trajectory.

Figure 9 illustrates again how the method of reciprocal values can unravel useful details about a studied process. Not only does it help in an unambiguous and easy identification of hyperbolic distributions but also it helps in an easy detection of deviations from such distributions. The world economic growth continues to increase but from the early 1900s it started to be diverted away from the faster accelerating historical hyperbolic trajectory to a slower trend.

The point of intersection of the reciprocal values with the horizontal axis is the point of singularity when the growth escapes to infinity. No growth can go beyond this point and any growth close to it may become unstable, unsustainable and catastrophic. Figures 8 and 9 show how close we are now to the point of the potential global economic instability and unsustainability.

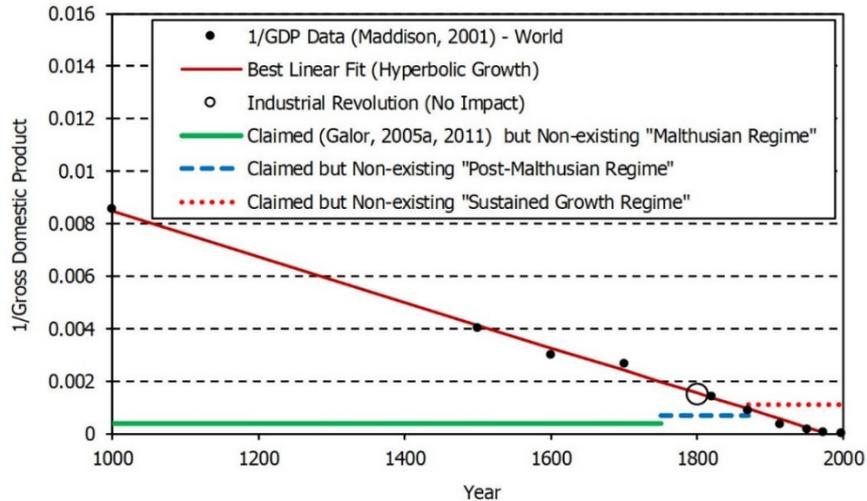

**Figure 8.** Fundamental concepts of the Unified Growth Theory are contradicted by *precisely the same data* (Maddison, 2001), which were used (but never analysed) during its development. Reciprocal values of data follow closely a decreasing linear distribution representing a monotonically increasing hyperbolic growth. The three regimes of growth claimed by Galor (2005a, 2011) did not exist. There was no takeoff around the time of the Industrial Revolution or around any other time. The monotonically increasing hyperbolic growth remained undisturbed until the 1990s. The GDP is in billions of the 1990 International Geary-Khamis dollars.

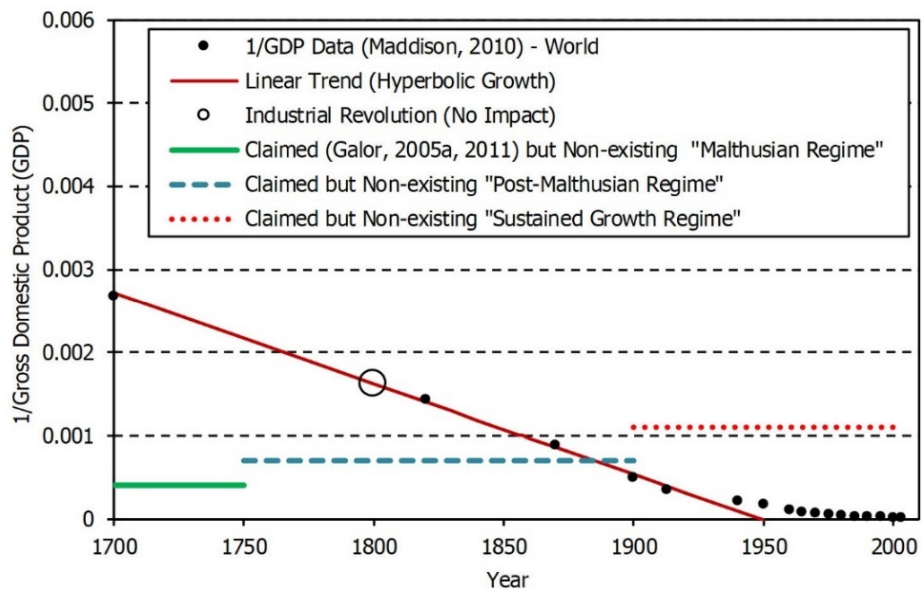

**Figure 9.** Maddison's data (Maddison, 2010) show clearly that while the Unified Growth Theory (Galor, 2005a, 2011) claims a transition from stagnation to a vigorous growth, the data show the opposite behaviour: a transition from a vigorous hyperbolic growth to a *slower* trajectory, as indicated by the upward bending of the growth trajectory of the reciprocal values during the 1990s and 2000s. There was no stagnation and no boosting in the economic growth at any time. The claimed (Galor, 2005a, 2011) but non-existent three regimes of growth are also shown. Their existence is contradicted by data. The GDP is in billions of the 1990 International Geary-Khamis dollars.

Unified Growth Theory claims that after a long epoch of stagnation we have now reached an era of "sustained economic growth," the term repeated 82 times in the first detailed formulation of this theory (Galor, 2005a), the potentially misleading description because while it is true that the current economic growth is still sustained the past economic growth was not only sustained but also it was increasing along a more secure trajectory, far away from the point of singularity. Even though the growth is now diverted to a slower trajectory any further increase can be potentially dangerous.

Reciprocal values of data show that for the first time during the AD era, and probably for the first time in human existence, we are now trapped between the already high level of economic growth and a point of no return, or equivalently between the very small reciprocal values of the GDP and zero. Any intrusion into this narrow gap has to be closely monitored. Even if the trend of the reciprocal values of the GDP data does not cross the horizontal axis any close approach to this axis could be dangerous, because it could trigger global economic instability and even a possible global economic collapse.

This simple analysis of data shows how dangerous are the generally accepted postulates presented in the Unified Growth Theory. The concept of a transition from stagnation to the so-called sustained growth regimes suggests that now for the first time in human history we can enjoy the sustained economic growth. Data, however, reveal a diametrically different pattern of growth. It was in the past that the economic growth was sustainable because it was following a stable hyperbolic distribution, which was still far away from the point of singularity. Now, however, the reciprocal values of the GDP are so close to zero, i.e. to the point when the GDP escapes to infinity, that the economic growth is no longer easily sustainable. The possibility of a serious economic instability is real. Now, the economic growth has to be closely monitored and controlled. The claim that we are now in the regime of sustained economic growth is dangerously inaccurate and misleading.

The method of reciprocal values can be also used to demonstrate that two other postulates of the Unified Growth Theory, the postulate of the differential takeoffs and the postulate of the great divergence, are contradicted by the mathematical analysis of data coming from *the same source,* which was used during the formulation of this theory (Nielsen, 2016b, 2016e, 2016i). Takeoffs never happened and consequently it makes no sense to claim that they happened at different times for developed and developing regions. The so called great divergence also never happened. Different regions are on different levels of development but they follow closely similar trajectories. They are like athletes running along similar tracks. They do not run in distinctly different directions as incorrectly claimed in the Unified Growth Theory but in the same direction.

If the economic growth continued along the historical hyperbolic trajectory it would have already reached a point of no return as indicated by the fitted straight line crossing the horizontal axis. To use the colourful description of von Foerster, Mora and Amiot (1960), we have been saved from experiencing a doomsday in the global economic growth. However, the danger of an excessive and unsustainable growth is still not averted.

Under a suitable control, the economic growth can continue for a long time, but this is precisely the important point: from now on the economic growth has to be closely monitored and controlled because it can easily become unsustainable.

Data between 1965 and 2003 follow closely exponential trajectory. Exponential growth does not increase to infinity at a fixed time but this is hardly any consolation because eventually such a growth also becomes unsustainable.

Any other continually increasing growth can be unsustainable unless it is increasing to a certain constant asymptotic value. However, it is extremely difficult to control such a growth because the growth rate would have to finely tuned to decrease slowly to zero. A constant growth rate, even if small, would represent the undesirable exponential growth. A growth rate fluctuating around zero would be safe but our general tendency is to try to increase the growth rate or at least to keep it constant, both options leading to unsustainable economic growth.

Data describing the world economic growth (Maddison, 2001) are compared in Figure 10 with the hyperbolic trajectory calculated using the straight-line fitted to the reciprocal values shown in Figure 8. Parameters describing the historical hyperbolic growth of the world GDP are: $a_0 = 1.716 \times 10^{-2}$ and $a_1 = -8.671 \times 10^{-6}$.

Now the puzzling features of the economic growth, the features that prompted so many discussions in numerous peer-reviewed scientific journals culminating in the formulation of the Unified Growth Theory, are manifestly clear, and their explanation is surprisingly simple. Over hundreds of years, the world economic growth was slow because it was hyperbolic. Over a short time, until the early 1900s, the economic growth was fast because it was hyperbolic – it followed *the same* undisturbed hyperbolic trajectory as in the past. The apparent transition from a slow to a fast growth is just an illusion created by the hyperbolic distribution. There was no unusually accelerated transition from the slow to the fast economic growth. The acceleration was gradual over the entire range of time.

The study presented here shows how important it is to have a clear understanding of the economic growth and how the simple method of reciprocal values can assist in such studies. Application of this method can not only assist in unravelling different components of growth trajectories but also to avoid being guided by hyperbolic illusions, which are the source of numerous misinterpretations of economic growth and of the growth of population culminating in the formulation of the fundamentally flawed and strongly misleading Unified Growth Theory (Galor, 2005a, 2011).

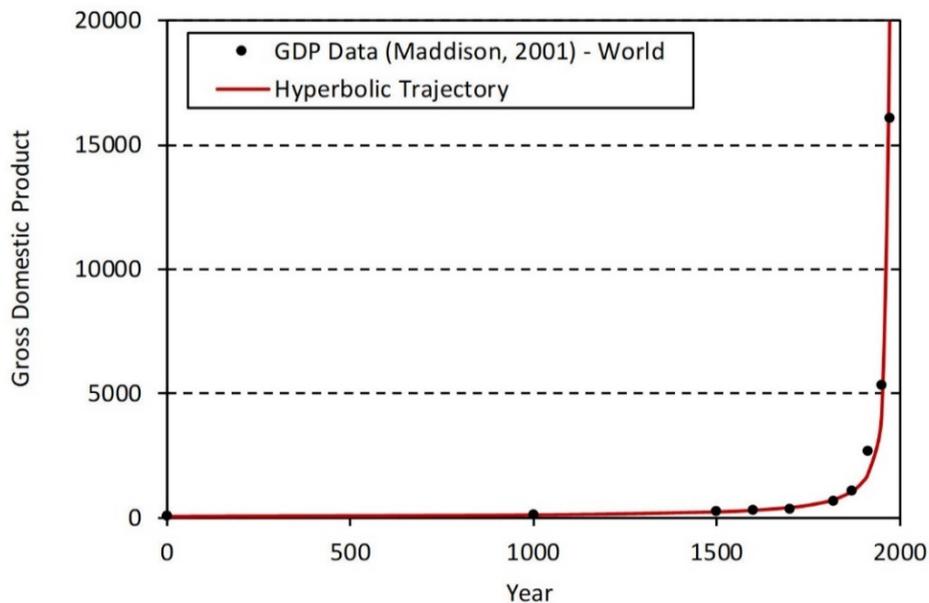

**Figure 10.** The data for the world Gross Domestic Product (GDP) (Maddison, 2001) follow closely the first-order hyperbolic distribution. The claimed three regimes of growth (Galor, 2005a, 2011) did not exist. They are replaced by an uninterrupted and monotonically increasing hyperbolic growth. The GDP is in billions of the 1990 International Geary-Khamis dollars.

## 4. Summary and conclusions

We have described a simple but effective method of analysis of hyperbolic distributions and we have explained its application by using the growth of the world population during the AD era. We have then demonstrated the flexibility of this method by using an example of the growth of human population in Africa. This method can be used to identify uniquely the first order hyperbolic distributions, to reveal hidden components of growth trajectories and to remove hyperbolic illusions, which are the source of numerous misinterpretations of economic growth and of the growth of population, misinterpretations prevailing over a long time in academic literature. This simple method redirects the economic and demographic research from explanations of phantom features created by hyperbolic illusions, to explanations based on the scientific analysis of data.

We have presented two other examples of analysis of data: the economic growth in Western Europe and the global economic growth. All these four examples show that the rigorous analysis of data contradicts the established knowledge in demography and in the economic research, and in particular, that it contradicts the fundamental postulates of the Unified Growth Theory (Galor, 2005a, 2011). However, what we have presented here is just a tip of an iceberg. An entirely new world is opened when we analyse more data (Nielsen, 2016a, 2016b, 2016c, 2016d, 2016e, 2016f, 2016g, 2016h, 2016i, 2016j, 2016k, 2016l, 2016m, 2016n, 2016o, 2016p, 2017a, 2017b, 2017c, 2017d), the world without stagnation in the economic growth and in the growth of population, without takeoffs from the alleged stagnation to growth, the world without complicated and untidy explanations of the mechanism of growth but the elegant world where data can be described by simple mathematical distributions, the world, which opens up new vistas for the demographic and economic research.

Impressions can be strongly deceptive and persuasive. "It is clear that the earth does not move, and that it does not lie elsewhere than at the centre" declared Aristotle. Fortunately, however, in science, incorrect interpretations are sooner or later corrected.

1. *Stagnation*. Research based on impressions and reinforced by the customary crude and self-misleading representations of data (Ashraf, 2009; Galor 2005a, 2005b, 2007, 2008a, 2008b, 2010, 2011, 2012a, 2012b; Galor & Moav, 2002; Snowdon & Galor, 2008)**,** such as shown in Figures 2 and 5, seems to confirm the generally accepted belief that there was an epoch of stagnation in the economic growth and in the growth of population. Scientific analysis of *precisely the same* (but undistorted) data demonstrates that there was no stagnation and that the economic growth and the growth of population followed *monotonically increasing hyperbolic distributions*.

2. *Takeoffs*. Research based on impressions seems to indicate that there was a transition from stagnation to growth described usually as a takeoff or explosion. Scientific analysis of precisely the same (but undistorted) data demonstrates that there was no takeoff or explosion and that economic growth and the growth of population continued to follow the monotonically increasing hyperbolic distributions. *What appears as a takeoff or explosion is in fact the natural continuation of hyperbolic growth.*

3. *Industrial Revolution*. Research based on impressions seems to indicate that Industrial Revolution played a crucial role in the economic growth and in the growth of population causing a dramatic acceleration (boosting) in the growth trajectories, described as takeoffs. Scientific analysis of precisely the same (but undistorted) data demonstrates that *Industrial Revolution had absolutely no impact on shaping growth trajectories*. Industrial Revolution can be linked to other impacts but not to shaping the population or the economic growth trajectories. This might be surprising but the evidence in data is undisputable and we have to accept it.

4. *Regimes of growth*. Research based on impressions seems to suggest that there were two or maybe even three distinctly different regimes of growth governed by distinctly different mechanisms (Galor, 2005a, 2011). Scientific analysis of precisely the same (but undistorted) data demonstrates that these two or three distinctly *different regimes of growth did not exist*. The growth was hyperbolic until recently when it started to be diverted to *slower* trajectories.

5. *Mysteries*. Research based on impressions resulted in claiming a series of "mind-boggling" and "perplexing" "mysteries of the growth process" (Galor, 2005a, pp. 177, 220). Scientific analysis of precisely the same data demonstrates that all these mysteries belong to the world of fiction created by a good dose of fantasy guided by the misleading impressions and reinforced by the customarily distorted presentations of data (Ashraf, 2009; Galor 2005a, 2005b, 2007, 2008a, 2008b, 2010, 2011, 2012a, 2012b; Galor & Moav, 2002; Snowdon & Galor, 2008) such as shown in Figures 2 and 5. Science is supported by a methodical analysis of data. There are no mysteries when precisely the same data are properly analysed.

    In particular, the mystery of the great divergence is explained: there was no great divergence (Nielsen, 2016i). Various regions are on different levels of economic growth but they all follow closely similar trajectories. Their economic growth did not diverge into distinctly different trajectories as incorrectly suggested by the crude representations of data.

    The mystery of the alleged sudden spike in the growth rate of income per capita has been explained: there was no sudden spike (Nielsen, 2016g). The growth rate of income per capita followed a monotonically increasing trajectory, which is readily represented by a mathematical distribution derived using hyperbolic growth for the growth of the GDP and for the growth of population.

    The mystery of the puzzling features of income per capita has been explained (Nielsen, 2017a). The distribution representing income per capita is nothing more than just a linearly modulated hyperbolic distribution. It reflects nothing more than the purely mathematical property of dividing two hyperbolic distributions.

    Other questions listed by Galor as representing the mysteries of the growth process can be easily answered. They refer to features that do not exist, features based on impressions reinforced by ineffectual handling of empirical evidence. They are in the same category as the question "Why does the sun revolve around the earth?"

6. *Mechanism*. Research based on impressions leads to proposing numerous complicated mechanisms of growth. Scientific analysis of data shows that the mechanism of growth is exceptionally *simple* (Nielsen, 2016p), which is hardly surprising because hyperbolic distributions are described by an exceptionally simple equation [see eqn (1)].

7. *Unified Growth Theory.* Research based on impressions prompted the development of a Unified Growth Theory (Galor, 2005a, 2011). Mathematical analysis shows that the fundamental postulates of this theory are *contradicted by the same data,* which were used during its development. Galor could have saved 20 years of his life and could have directed his academic skills to developing a useful theory if he did what any scientist is supposed to do: if he based his deductions and explanations on a scientific analysis of data. He had access to excellent data but he did not analyse them. He was guided by preconceived ideas and he supported them by distorted presentations of data.

The analysis data suggests new lines of research. Thus, for instance, the relevant question is not why the historical economic growth was so unstable in the past or what caused the perceived transition from alleged stagnation to growth but *why the economic growth was so remarkably stable* in the past. The same question applies to the growth of population but it was already answered (Nielsen, 2016c, 2017d). The growth of population was remarkably stable because of the combination of the generally low impacts of demographic catastrophes (at least on the global and regional scales) and the high level of human resilience expressed in the efficient process of regeneration (Malthus, 1798; Nielsen, 2016c). If we accept that there is a close relationship between the growth of population and the economic growth, then the question about the stability of the historical economic growth has been also already answered. However, it is possible that some new insights could be still added to this explanation.

The relevant question is not why the Industrial Revolution and the unprecedented technological development boosted the economic growth because they did not. The relevant question is *why the Industrial Revolution and the unprecedented technological development did not boost the economic growth*. Why these apparently strong technological and socio-economic forces had no impact on shaping the economic growth trajectories.

The relevant question is not why the economic growth increased so fast in modern time, because we have shown that this fast increase was just the natural continuation of the monotonically increasing hyperbolic growth until in recent years it started to be diverted to a *slower* but still fast-increasing trajectory. The relevant question is *why the economic growth was diverted to a slower trajectory*. What new force or forces were so strong that they were able to overpower the historically strong force of growth. Another relevant question is also whether this new trajectory is likely to develop into a historically preferable and potentially catastrophic, hyperbolic growth. Furthermore, the relevant question is how to control the current fast economic growth. The same question applies also to the growth of population but it was at least partly answered in the study of the effects of Malthusian positive checks (Nielsen, 2016c). The primary if not exclusive way of controlling the growth of human population is to improve the living conditions in developing countries.

The method of reciprocal values is so simple that it can be used by anyone and it is, therefore, expected that it will be of interest to many scientists who look for a simple method of analysis of empirical evidence, a method that does not involve any complicated mathematical formulae, any intricate mathematical algorisms or the use of powerful computers but a simple display of data and a remarkably simple fitting procedure. We have demonstrated that even a simple mathematical method can have a dramatic influence on scientific research.

It is essential to understand that by claiming that there was no stagnation in the economic growth or in the growth of population we are not claiming that there was no stagnation in the standard of living. We are only claiming that the two processes were decoupled. We might, if we insist, describe the past general living conditions as primitive or even stagnant, but there is no evidence that they were shaping the trajectories describing the growth of population or the economic growth.

It is also essential to understand that by claiming that there was no takeoff in the economic growth or in the growth of population we are not claiming that there was no takeoff in the technological development, or generally in the intellectual progress and in the dramatic changes in human experience and in living conditions. We are only claiming that these possible takeoffs had no impact on changing the economic growth trajectories or the trajectories describing the growth of population. There were no takeoffs in any of these two processes. Industrial Revolution can be linked with many changes in human living experience but all these changes had no impact on changing the economic or demographic growth trajectories.

There is no reason why scientific evidence presented here and in other related publications (Nielsen, 2016a, 2016b, 2016c, 2016d, 2016e, 2016f, 2016g, 2016h, 2016i, 2016j, 2016k, 2016l, 2016m, 2016n,

2016o, 2016p, 2017a, 2017b, 2017c, 2017d) should not be accepted by the scientific community. The only alternative option is to reject data but this would be no longer science.

Even Galor and his associates accept the same data and use them in their research. Their unfortunate mistake was only in choosing to support their investigations by the grossly distorted and self-misleading representations of data (Ashraf, 2009; Galor 2005a, 2005b, 2007, 2008a, 2008b, 2010, 2011, 2012a, 2012b; Galor & Moav, 2002; Snowdon & Galor, 2008). Consequently, the only way to reject scientific evidence and to accept the doctrines of stagnation and takeoffs, and all other associated erroneous explanations of the dynamics of the economic growth and of the growth of population, is to accept data but distort them in such a way as to make them to conform with preconceived ideas, but then again it is not science.

Evidence in data is overwhelming and leaves no room for accepting incorrect interpretations. In order to have progress in the demographic and economic research, incorrect interpretations of growth have to be abandoned and a new paradigm has to be developed. There is no other, scientifically justified, way. A serious mistake in scientific investigations is not in stumbling and in making mistakes but in refusing to learn from them and to correct them.

Pólya, G. (1981). *Mathematical Discovery: On Understanding, Learning, and Teaching Problem Solving*. John Wiley & Sons, Inc., Hoboken, NJ.

Robine, J-M. (2001). Redefining the Stages of the Epidemiological Transition by a Study of the Dispersion of Life Spans: The Case of France. *Population*, *13*(1), 173-193.

Smil, V. (1999). Detonator of the Population Explosion. *Nature*, 400, 415. http://dx.doi.org/10.1038/22672

Snowdon, B., & Galor, O. (2008). Towards a Unified Theory of Economic Growth. *World economics, 9*(2), 97-151.

Steinmann, G., Prskawetz, A., & Feichtinger, G. (1998). A model on the escape from the Malthusian trap. *J Popul Econ, 11*(4), 535-550. http://dx.doi.org/10.1007/s001480050083 PMid:12294786

Strulik, H. (1997). Learning-by-doing: Population pressure, and the theory of demographic transition. *Journal of Population Economics, 10*(3), 285-298. http://dx.doi.org/10.1007/s001480050044 PMid:12292961

Tamura, R. F. (2002). Human capital and the switch from agriculture to industry. *Journal of Economic Dynamics and Control, 27*, 207-242. http://dx.doi.org/10.1016/S0165-1889(01)00032-X

Thomlinson, R. (1965). *Population dynamics: Causes and consequences of world demographic change.* New York: Random House.

van de Kaa, D. J. (2008). Demographic Transition. *Encyclopedia of Life Support Systems, Demography, Vol. 2*. Paris: UNESCO

Voigtländer, N., & Voth, H.-J. (2006). Why England? Demographic factors, structural change and physical capital accumulation during the industrial revolution. *Journal of Economic Growth*, *11*(4), 319–361. http://dx.doi.org/10.1007/s10887-006-9007-6

Vollrath, C. (2011). The agricultural basis of comparative development. *J Econ Growth, 16*, 343–370. http://dx.doi.org/10.1007/s10887-011-9074-1

von Foerster, H., Mora, P., & Amiot, L. (1960). Doomsday: Friday, 13 November, A.D. 2026. *Science, 132*, 255-296. http://dx.doi.org/10.1126/science.132.3436.1291

Wang, C. (2005). Institutions, demographic transition, and Industrial Revolution: A Unified Theory. SSRN Working Paper Series, 08/2005.

Warf, B. (2010). Demographic transition. In B. Warf (Ed.), *Encyclopedia of Geography (pp. 708-711)* Thousand Oaks, CA: *SAGE*.

Weisdorf, J. L. (2004). From stagnation to growth: Revising three historical regimes. *Journal of Population Economics*, *17*, 455-472. http://dx.doi.org/10.1007/s00148-004-0182-5

Weiss, V. (2007). The population cycle drives human history – from a eugenic phase into a dysgenic phase and eventual collapse. *The Journal of Social, Political, and Economic Studies*, *32*(3): 327-358.